\documentclass[12pt,preprint]{aastex}

\shorttitle{Linking Literature and Data}
\shortauthors{Accomazzi}

\begin{document}

\title{Linking Literature and Data: Status Report and Future Efforts}
\author{Alberto Accomazzi}
\affil{Harvard-Smithsonian Center for Astrophysics, 60 Garden Street, Cambridge, MA, 02138 USA}
\email{aaccomazzi@cfa.harvard.edu}

\begin{abstract}
In the current era of data-intensive science, it is increasingly
important for researchers to be able to have access to published 
results, the supporting data, and the processes used to produce
them.  Six years ago, recognizing this need, the American Astronomical
Society and the Astrophysics Data Centers Executive Committee (ADEC)
sponsored an effort to facilitate the annotation and linking of
datasets during the publishing process, with limited success.  I will
review the status of this effort and describe a new, more general
one now being considered in the context of the Virtual Astronomical
Observatory.
\end{abstract}

\section{Introduction}

Links between papers in ADS and data products hosted by astronomy
archives have existed since 1995.  These links have been created
and curated by librarians and archivists as part of the 
data center's effort to collect information about the scientific
use of the data being hosted by the archive.  The links provide more than just
useful connections between bibliographic records and observations
that allow users to access related material.  They 
represent part of the scientific artifacts created during the 
research lifecycle of an astronomer, and as such are needed
to fully document and describe the research activity itself
\citep{LISAVI}.

One obvious benefit which comes from maintaining such links is
the ease with which one can generate metrics about
the scientific impact of the observations in the form of
published papers or citations.
Thus, in a highly competitive scientific discipline such
as astronomy, maintaining the complete record of paper-data
connections has become an accepted way to evaluate a
project, mission, or even an entire research field
(\cite{Uta}, \cite{2010AN....331..338T}).
The metadata collected during the creation of 
links maintained by ADS and its collaborators have so far been
limited to some very basic information about the location
of the resources linked together.  Typically these are
simple mappings of ADS's bibliographic identifiers 
({\it bibcodes}) and URLs pointing to one or more 
particular data product(s) hosted by an archive.  
The ADS record will of course
also have a link to the published paper itself, acting as a
``bridge'' between the manuscript and the data described
therein.  Figure 1 shows the connection between a paper and
data products available from the Multimission Archive at
Space Telescope (MAST) and the Chandra X-Ray Archive.

\begin{figure}[!ht]
\includegraphics[width=1.0\textwidth]{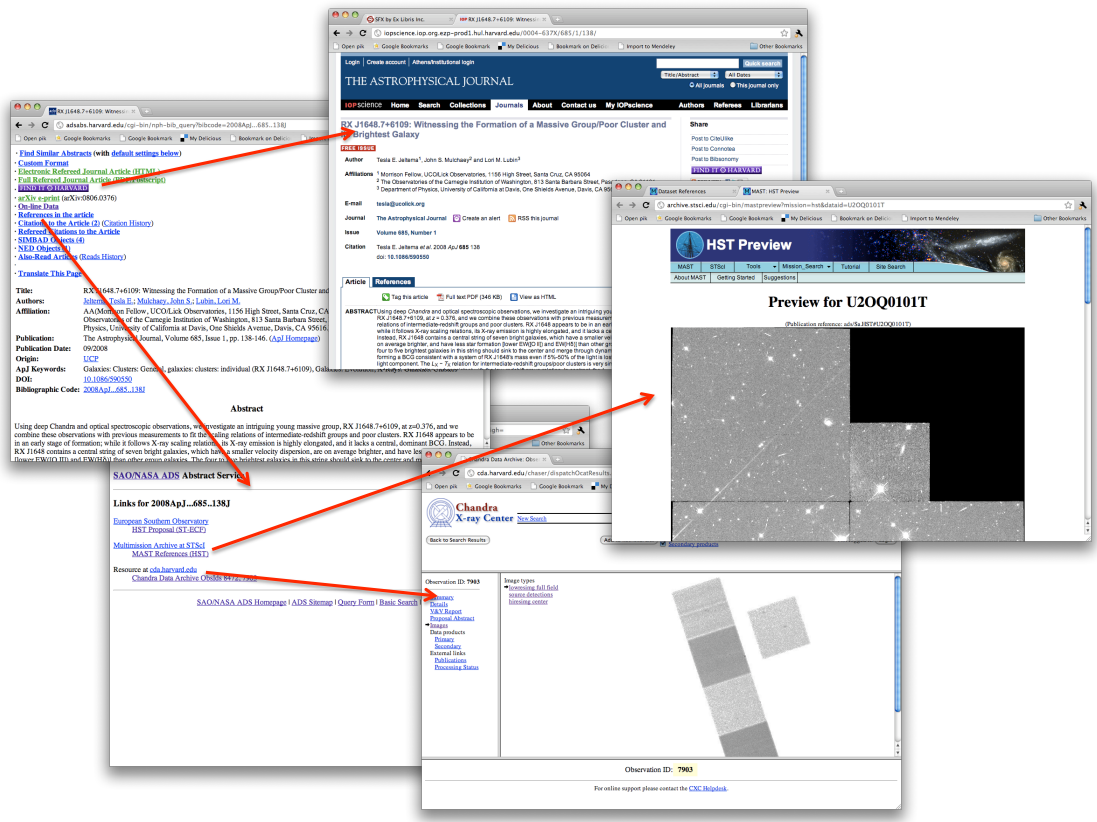} 
\caption{Links between an ADS record, the full-text manuscript
hosted by a publisher, and data products available from
MAST and Chandra}
\label{fig1} 
\end{figure}

\section{Development of Dataset Linking Infrastructure}

In 2002, it became apparent
that the methodology adopted to create, maintain and share 
these linkages could be improved.  Thus, in 2003, the NASA
Astrophysics Data Centers Executive Council (ADEC) and the
AAS journals issued guidelines aimed at improving the
situation \citep{2004AAS...204.7502E}.  
This new effort was aimed at addressing
four separate issues in the management of these links: 
their curation, naming, resolution and persistence.

The creation of links to data products has been a time-consuming
activity usually carried out by a librarian or archivist.
\cite{2004ASPC..314..605R} describe the effort required to perform this
activity, which typically consists of scanning the literature
to identify which papers mention one or more data products
from a particular archive, and then link those papers with the 
relevant datasets.

In 2004, in order to facilitate this activity, and in coordination
with the ADEC proposal,
the Astrophysical Journal introduced the capability for authors
to properly tag the datasets analyzed in the paper.  This 
introduced a mechanism to formally ``cite'' data in a way similar 
to how scientists cite other papers.  
According to this plan, citations to data products would be vetted
by both editors and referees during the manuscript editorial process,
and links would be created to the corresponding data
products as part of the process which generates the online HTML
version of the paper.  The correlation between a paper and the
datasets referenced therein would then be propagated back to 
the ADS and the participating data centers via metadata 
exchange.

The implementation of this linking proposal would not only benefit 
end-users, but would potentially provide significant savings in the 
curation efforts of archivists and librarians, who could now
harvest these linkages directly from ADS, thus reducing the need
for the manual scanning of the literature.

In order to properly cite the datasets in the literature, the
ADEC and AAS adopted a standard way to uniquely identify data
resources based on the IVOA Identifier standard 
\citep{IVOAIDS}.  The proposed system of nomenclature
\citep{2007ASPC..376..467A} provided a standard for dataset
identifiers which featured some important properties.  Among
them: uniqueness (one resource corresponds to a single 
identifier), and persistence (identifiers do not change even when
data products are migrated to a different archive).
The identifiers were designed to support the naming of resources 
with a broad range of granularity and included a ``public'' 
prefix identifying the archive or mission that generated the dataset
as well as a ``private'' key identifying the data item within
a specific collection.

In order to ensure the proper use and persistence of links to datasets,
the ADEC charged the ADS with the task of setting up a verification and
resolution service for dataset identifiers.  In this role, the ADS would
act as the registration authority on behalf of the community, creating
the infrastructure necessary to enable the dataset linking.
During copy-editing of a paper, the editors would use an automated 
tool provided by ADS to verify that a particular dataset identifier
is known and can be resolved to an online resource.  Upon successful
verification, the identifier would be incorporated into the paper with
a link to a resolution service provided by ADS (rather than a simple
link to the current URL for the resource).  This model provides a level
of redirection which can be used to properly track a dataset if and
when it moves from one archive to another, and allows the resolver to
provide options should multiple versions of a data product be available.
A complete description of this implementation can be found in \cite{2007ASPC..376..467A}.
Elements of this architecture are similar to the Digital
Object Identifier standard used for the persistent linking of scholarly 
publications, which is discussed in section 4.
However,  this system was designed to be fully managed by the astronomical
community requiring a minimal level of effort for institutional buy-in.

\section{Current Status}

Six years have passed since the introduction of the dataset 
linking infrastructure, so now is a good time to take stock of
this effort.  
From a system design point of view, 
some of the features that made the implementation 
of this system attractive have, in retrospect, proven to be obstacles to its
long-term success.  Chief among all problems with the registration
of dataset identifiers has been enforcing their persistence.  
Since data products ultimately reside within archives that participate in the ADEC but which
are run independently of each other, the implementation and maintenance of
services that provide access to the data is left to the 
archives themselves.  Given that the thrust of this effort is completely
voluntary, there is no contract or reward system which
can be leveraged to enforce the long-term
resolution of and access to a particular dataset. 
Experience shows that unless 
requirements for the preservation of these linking services become
part of the archive operations, a simple system upgrade is enough
to break valuable links to dataset resources.
As an example, over 200 dataset identifiers which were published in
a 2004 ApJ Supplement special issue on the Spitzer Space Telescope
are no longer resolvable due to a change in the Spitzer 
Science Center interface.

\begin{figure}[!ht]
\includegraphics[width=1.0\textwidth]{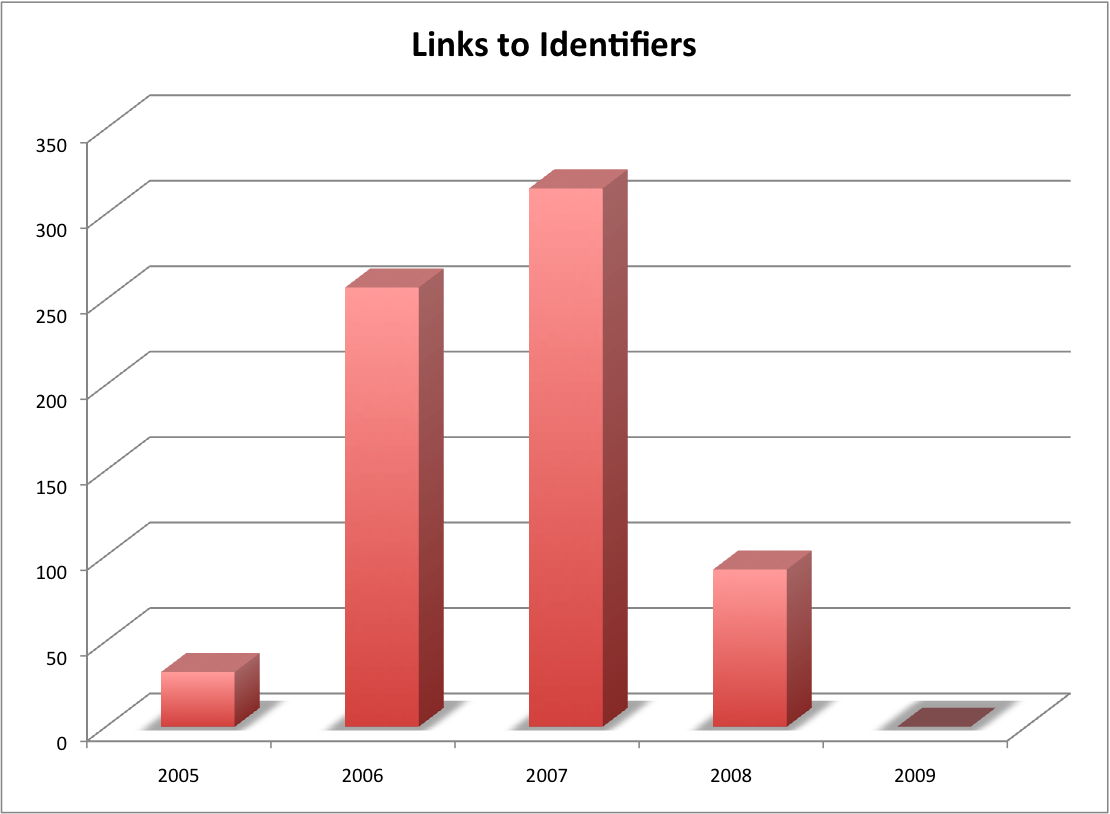} 
\caption{The number of links to data entered in AJ and ApJ papers
as part of the ADEC/AAS dataset linking effort}
\label{fig2} 
\end{figure} 

Unfortunately the adoption and use of dataset identifiers in the 
literature has not been a success story.  Citations
to datasets began appearing in 2005 and increased in the following two years, peaking
in 2007, before decreasing in 2008 and finally going down to zero in 2009
(see fig. 2).
The reasons for this reversal are not entirely clear, but can
probably be attributed to a variety of factors.  First and
foremost, even though the ADEC approved a policy encouraging archives
and users to make an effort to more widely publish dataset identifiers,
anecdotal evidence shows a low level of awareness 
from scientists of this possibility.  
Researchers tend to be busy and, unless properly coached by editors
and archivists, will easily overlook new demands or stipulations
requiring additional work on their part.
In addition, data archives don't always make it 
obvious how a particular dataset
(or data file) should be cited in the literature by the scientists
who make use of the data in their research.  Since astronomers have
been accustomed to referring to the data they have used in terms of 
specific observations or regions of the sky, this general practice is still
used.
Rather than unambiguously identifying the data using dataset identifiers,
astronomers describe how the data can be obtained.
While this is a reasonable way for an author to convey the necessary
information about the data being studied, it obviously defeats the
goal of creating persistent, unambiguous, machine-readable links to 
the data products.
Finally, it was hoped that after the initial adoption of the standard 
for dataset identifiers within the ADEC more data centers 
and more journals would follow suit, but this did not materialize.  
The critical mass and community awareness necessary to make this 
common practice was never reached.
 
\section{A Way Forward}

Despite the lack of adoption from the community of the proposal
described above, every scientist, librarian and archivist
agrees that preserving data products and publishing links to data
in the literature is a worthwhile effort.  
We believe that at this point in time we should even be
more ambitious, and recognize that in fact for our discipline to
flourish in the digital era all artifacts related to the research
lifecycle need to be available online, and properly interlinked
\citep{LISAVI}.  Thus, the issue of creating links from the literature
to data products can be recast in a wider scope -- the
preservation and interlinking of digital assets in astronomy.
The use of the term ``digital assets'' in this context refers to
artifacts used and generated during the research
activity of an astronomer.
This includes observing proposals, observations, archival data from 
surveys and catalogs, observing logs, 
tables and plots that are published in manuscripts. 
In short, we advocate capturing all the data and knowledge that has
gone into the research activity itself, with the aim of
providing a digital environment that can support the
repeatability of the research described in a publication.
While there are many possible implementations of a digital
environment for data preservation, it is clear that any
such effort must satisfy a set of principles.
Below we identify some of the basic requirements that we
believe will need to be addressed in the near future.

\subsection{Management of Digital Assets}

First, one should consider the issue of nomenclature and
persistence of digital assets.
Since we can expect that the data referred to in a paper
will be hosted on a distributed set of digital repositories,
naming and linking standards need to be clearly defined and adopted  
in order to create persistent links to such resources .
The solution proposed by the ADEC 
was primarily designed to satisfy the 
requirements of uniqueness and persistence 
for data already available in well-established archives
using community-developed
standards.  While this approach is technically sound and seemed
at the outset to provide the best solution to the problem,
new technologies and standards since developed by the digital library 
community now offer attractive alternatives that should be considered.
For the creation and management of unique identifiers, 
the Handle System is a general purpose distributed
information system for the minting and resolution of unique
identifiers on the internet.  
The Digital Object Identifier (DOI) system is an application
built upon the Handle system and is
widely used by the digital
publishing industry.  Organizations making use of the DOI system 
agree to a business model that requires the deposit and active
curation of metadata for the digital assets registered in the system,
and are subject to fines if found to be in breach of the DOI Foundation
rules.  This elevated level of commitment provides a certain level of assurance that the
digital assets registered in the system will be properly maintained.
In addition, the DOI foundation explicitly imposes requirements
on its members to ensure the long-term survival of the system.  For instance,
should one of its members cease operations, the DOI resolution
of its content would be transferred to other members of the
foundation.  
When it comes to the preservation of data products, 
this type of long-term commitment has never
been formalized or made explicit by most of the 
publishers, societies or even astronomical data centers
(except for the case of the active NASA missions, whose digital
assets are transferred to archival centers at the end of a mission).
However, it is exactly the type of commitment we believe
is essential for our community to make at this time.

\subsection{Archival and Preservation}

In order to enable the publication and 
broader re-use of scientific high-level data products,
researchers should be required to upload
such data to one or more trusted, community-curated, digital 
repositories.  Not only does this requirement allow
repeatability of experiment and analysis, but it promotes
a level of transparency and trust that is an important
component of the scientific discourse.

While much of the tabular data now published
in scientific articles ends up being stored in services
such as NED and Vizier, a significant amount of supplementary
material does not make it into such archives.  
To be sure, authors are often encouraged to submit
machine-readable versions of the data (or even computer
code) as supplementary material submitted to the journal,
but the uniformity, longevity,
re-usability and discoverability of such material are 
at this point highly inconsistent and questionable.  In addition, no explicit
or common migration plan has so far been defined by 
publishers or learned societies, so future access and curation of 
these assets is not assured.
We believe that it is essential to encourage the
deposit of digital assets in a wide range of trusted repositories 
curated by the community in collaboration with the journals.
The kind of material deposited in such repositories will
supplement the products which are currently curated by 
projects such as NED and Vizier, and consist
of anything which does not fit in the typical description
of a data table or catalog.  This may include 
published images, plots,
observing notes, workflows, software,
intermediate results, and large data collections.
In order for these data products to be useful, it is
essential for the user to deposit and for the repository to
expose not just raw data but also its related metadata.
This should include, at a minimum, a description of the datasets 
in the sample, a set of applicable keywords, and some
notes relating the data in question to the published paper(s)
in which they were used.

The need to create a digital infrastructure in support of these
activities is increasingly being recognized both in the US and in Europe
by funding agencies such as  NASA, the NSF, JISC and digital
preservation programs are now being defined \citep{Sayeed}.
In particular, the NSF DataNet Data
Conservancy program has established as one of its goals
the support of scientific inquiry through the adoption of
a comprehensive data curation strategy.
Today there are a number of open-source digital repository
systems available, some of which have already been deployed
by universities and projects involved in preserving 
digital institutional assets (e.g. Fedora or DSpace).
Other initiatives, such as the Dataverse Network
\citep{King},  provide a scientist-centered framework for storing 
data products associated with publications and encourage their citation 
through the use of unique, permanent dataset identifiers
based on the Handle system.

\section{Discussion}

Even though there is general agreement that publishing and
citing data 
is a noble goal and worthwhile effort, the experience of the ADEC
data linking effort has shown that it takes sustained community
engagement to turn a proposal into a successful activity.
When many people and organizations are involved in
providing crucial components of such a distributed system, the risk
of multiple points of failure becomes significant and can ultimately
spell the demise of even the best thought-out technical scenario.
Rather than giving up on this worthwhile idea, 
we should take the failure of adoption as a learning opportunity to devise
a more robust system that can not only provide links to existing
datasets available from well-established archives, but 
also provide the capability of storing author-supplied
data and metadata related to publications.

In the long run, it is likely that there will be an ecosystem of
different repositories and technologies used for the 
preservation of research products.  Some of them will be more
focused on actively capturing and curating datasets, as is today
done by projects such as Vizier and NED, and others which will provide
an infrastructure which can be used by scientists and publishers to
self-manage data products associated with published papers.
The overarching goal of such systems should be to provide useful
services to the astronomy community and guarantee that the data
deposited in such repositories will be properly curated and preserved
for the foreseeable future.  This includes providing a migration
path for obsolete data formats, curating and exposing metadata
of digital assets in the repository, and providing 
discovery services to its content.
This commitment comes at a cost, which
should be shared by the community as part of the effort which funds
the infrastructure supporting astronomical research.
In addition, it is
essential that we promote and encourage policies that
foster and facilitate the growth of the digital scholarly environment
that the Virtual Observatory has been envisioning.
The recent funding of digital preservation frameworks such as
the Data Conservancy project suggests that the time has come for
the VO to play a major role in the capture and preservation of
the astronomy research lifecycle.  
We look forward for the
members of the International Virtual Observatory Alliance to 
take a pro-active role over the next decade in order to make this
vision a reality.

\end{document}